# Spatial and temporal analysis of political violence in the United States

Ravi Varma Pakalapati and Gary E. Davis

Department of Mathematics, University of Massachusetts Dartmouth, USA

**Abstract**

Acts of political violence in the continental United States have increased dramatically in the last decade. For this rise in political violence, we are interested in where and when such incidents occur: how are the locations and times of incidents of political violence distributed across the continental United States, and what can we learn from a detailed examination of these distributions? We find the distribution of locations of political violence is neither uniform nor Poisson random, and that such locations cluster into well-defined geographic regions. Focusing on the county level we find a markedly skewed distribution of county counts of incidents of political violence. Examination of news reports and commentaries provided by the Armed Conflict Location & Event Data Project for the extreme outlier counties reveals compelling political and social background to the reported incidents of political violence. This, together with credible information on the role of social media in fomenting political violence leads us to postulate a field notion of upsetness as a major background to political violence. Using the time stamp on incidents of political violence we constructed a nearest neighbor model to predict future incidents of political violence at specific locations that involved a fatality.

## 1. Introduction

We examine the spatial and temporal distribution of incidents of political violence in the continental United States, 2020 through 2024. The term "political violence" is used by the Armed Conflict Location & Event Data Project (ACLED) to categorize certain forms of demonstration and protest. This term has been analyzed by Kalyvas, 2019, who details eleven varieties of political violence: interstate war, civil war, state repression, genocide, ethnic cleansing, intercommunal violence, organized crime/cartel violence, military coup, mass protest/rebellion, political assassination, and terrorism. In this article we deal with ACLED data categorized as political violence that generally fits into the Kalyvas categories of *intercommunal violence* and *mass protest*.

Acts of political violence in the United States have increased dramatically in the last decade (ref. Kleinfeld, 2021, p. 160). For this rise in political violence, we are interested in where and when such incidents occur: how are the locations and times of incidents of political violence distributed across the continental United States, and what can we learn from a detailed examination of these distributions?

This article is structured as follows: first we examine the characteristics of the geographic distribution of unique locations of political violence across the continental United States, noting the randomness or otherwise of those locations. We then use an unsupervised clustering technique to ascertain whether the locations of political violence fall into distinct geographic clusters. An uneven geographic distribution of locations of political violence stimulated us to examine political violence at the *county level* - an administrative unit below that of a state. From the distribution of county level political violence, we identify extreme outliers, and for those outliers we extract word clouds from associated news reports at the time of the incidents. About one in ten incidents of political violence involve one or more fatalities. We examine connections between number of incidents of political violence per county and county population to address whether there is a correlation between incidents of political violence and population. We consider a nearest-neighbor model to predict future incidents of political violence, by time and location, in which there is at least one fatality, as a forecasting method for law enforcement. Details of the political and social situations at certain locations of political violence lead us to consider a *field* notion of upsetness (in the sense of Martin, 2003). Finally, we reflect on the increasing influence of social media on acts of political violence both in the United States and other jurisdictions.

## 2. Data and methods

The data for this study came from the Armed Conflict Location & Event Data Project (ACLED), from 2020 through 2024. ACLED provides comprehensive information on political violence and protest events worldwide, including dates, locations, actors involved, fatalities, and event types:

> "The Armed Conflict Location & Event Data Project (ACLED) is a disaggregated data collection, analysis, and crisis mapping project. ACLED collects information on the dates, actors, locations, fatalities, and types of all re- ported political violence and protest events around the world." (About ACLED, https://acleddata.com/about-acled/)

While the ACLED data contains information on various types of protest events we choose to focus solely on events categorized as involving some form of politically motivated violence ("political violence" for short). The ACLED data provides evidence for the occurrence of political violence both by location and time, and so provides background data that informs critiques of models for the rise of politically violent actors.

The principal tool we used for analyzing the data on political violence and models of violent actors was *Mathematica* (ref. Dauphine, 2017). *Mathematica*'s extensive geo-computation functions and capacity to access the Wolfram Language geo-computation tools proved invaluable, especially for spatial and temporal analysis of the political violence data.

*Mathematica* tools and functions that proved especially useful were: GeoPosition, GeoListPlot, GeoHistogram, GeoSmoothHistogram, GeoDistance, GeoIdentify, GeoEntities, GeoPath, FindClusters (KMeans). We utilized *Mathematica*'s computational ability to assign a US county to latitude/longitude data through its GeoIdentify function.

We also utilized Python's Natural Language Toolkit to construct word clouds from news reports recorded in the ACLED data.

## 3. Results

### *3.1 Spatial distribution*

We constructed a geographic visualization of the unique locations of occurrences of political violence in the continental USA (January 1, 2020, through August 1, 2024) Utilizing Mathematica's GeoPosition and GeoListPlot functions, as shown Fig. 1.

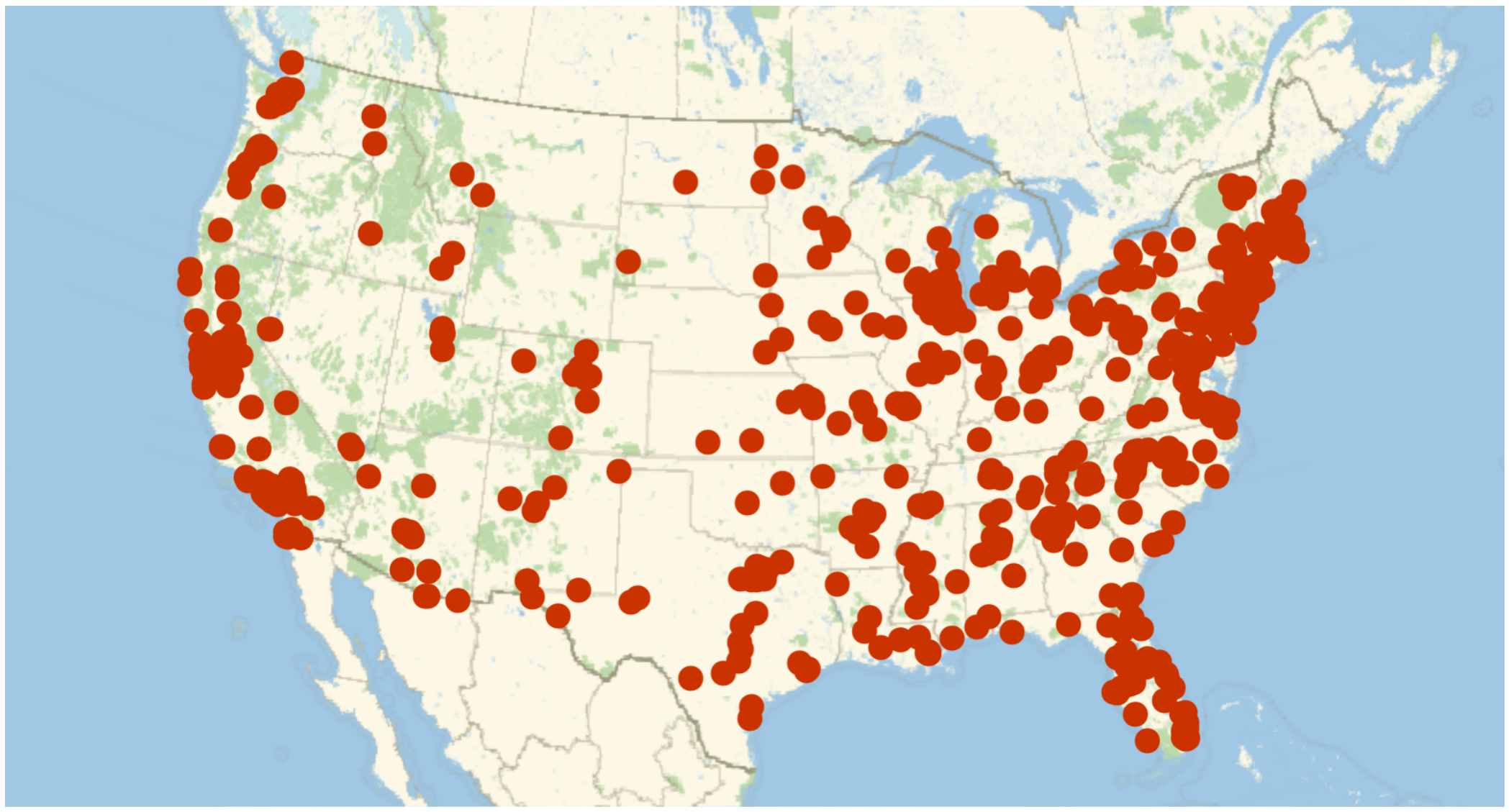

Fig. 1. Unique locations of incidents of political violence, USA

We can see from the smooth approximation to the density of the locations that the unique locations appear to be neither uniform random nor Poisson random, as indicated in Fig. 2.

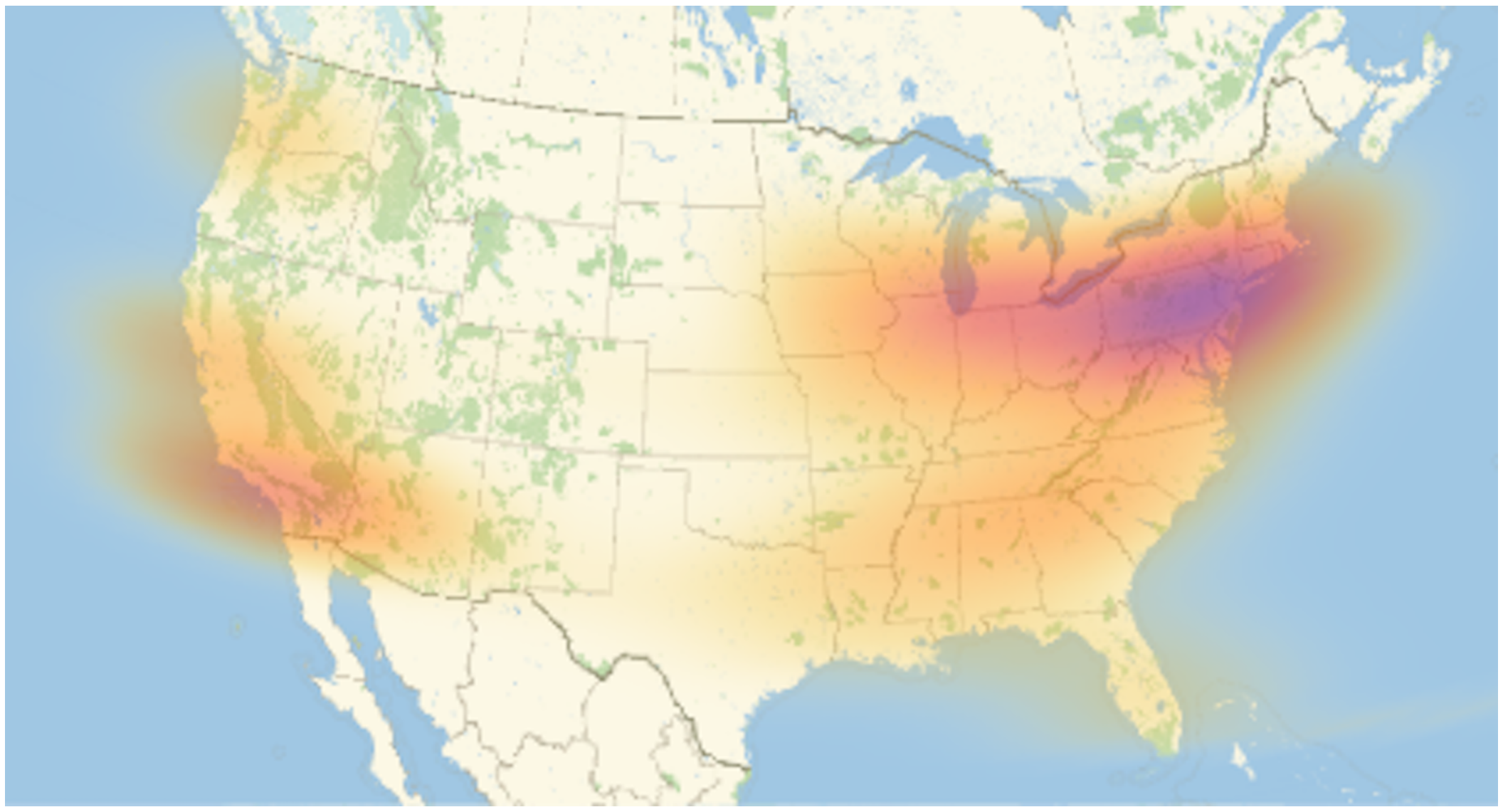

Fig. 2. Smooth histogram of locations of incidents of political violence

Recall (Kingman, 1992) that for a homogenous Poisson point process:

1. The underlying rate at which locations of political violence occur doesn't change with position in the continental US (homogeneity).
2. The number of locations of political violence in separate regions are independent.
3. No two locations of political violence can be in exactly the same place.

The 3rd condition is trivially true because we are listing only unique locations (not how often a politically violent act occurs at that location). The second condition suggests a strong reason why the locations do not follow a homogenous Poisson point process, since it is highly unlikely that acts of violence occurring in different regions of the US, for example Portland Oregon and Los Angeles, California, are independent, since acts of political violence are often, if not usually, driven nowadays by widely broadcast news events or social media.

We previously used several analytical tests for randomness, both uniform and Poisson, which indicated convincingly, as the images above suggest, that the locations of incidents of political violence are far from random in the continental US (ref. Pakalapati, 2025). As an example of an analytical test for uniform randomness we mention distance to a nearest neighbor as a measure of spatial homogeneity (Clark & Evans, 1954). For the 600 unique locations of political violence in the continental US we calculate a nearest neighbor. The distribution of nearest neighbor distances is shown in Fig. 3.

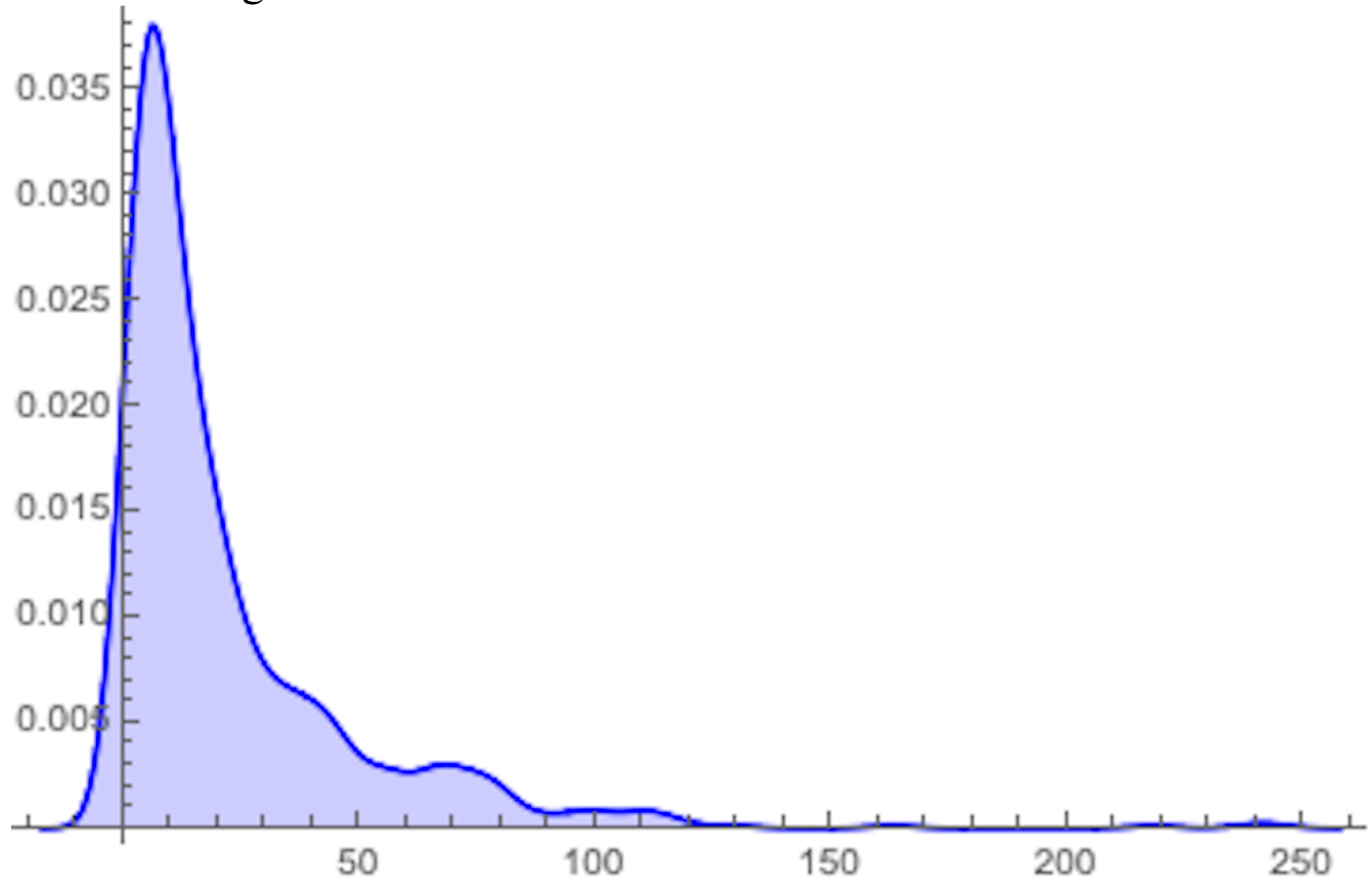

Fig. 3. Nearest neighbor distribution, US locations of political violence

We compare this with the nearest neighbor distribution of 600 uniformly random locations in Fig. 4.

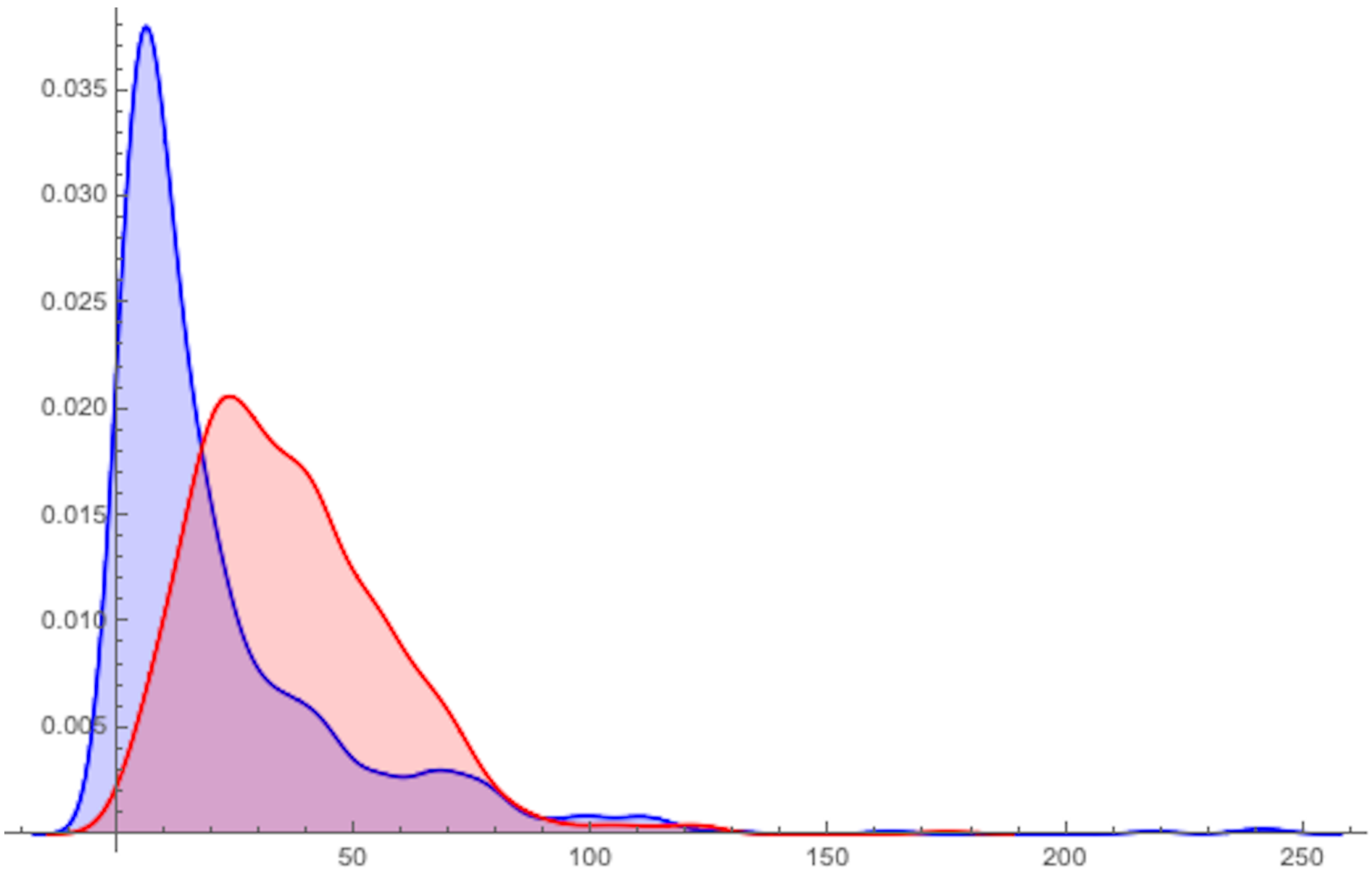

Fig. 4. Nearest neighbor distribution, US locations of political violence together with nearest neighbor distribution, 600 uniform random US locations

A highly significant difference in the locations of political violence and the uniform random locations is in the *variance* of the nearest neighbor distributions. For the locations of political violence, the variance is 824.18, while for the uniform random locations the variance is 418.76. The theoretical variance for uniform random points, given density $\rho = 600/3706269$ is $(4 - \pi)/(4\pi\rho) = 421.96$ (ref. De Smith, Goodchild & Longley, 2007), close to the variance for the empirical uniform random distribution of locations. We conclude that the locations of incidents of political violence in the continental US are far from uniformly random.

*3.2 Spatial clustering*

Since we know the locations of political violence in the continental US are not random, we examine to what extent the locations fall into well-defined clusters, and what is the nature of those clusters. The non-randomness of the locations indicates there is some structure in the data and we use k-means clustering to begin to address the nature of such structure.

k-means clustering of the locations of political violence aims to partition the locations into k clusters in which each location belongs to the cluster with the nearest mean, where "nearest" refers to geographic distance on the Earth's surface. k-means clustering is an unsupervised learning technique to classify unlabeled data, and as such reduces human bias as to the nature of clusters (ref. James et al., 2023). We produce k-means clusters for k = 2, 3 and 4. As shown in Fig. 5, for k = 2, k-means clustering provides a distinct North-South continental divide in the locations of incidents of political violence, with 314 incidents in the North cluster and 286 in the South cluster:

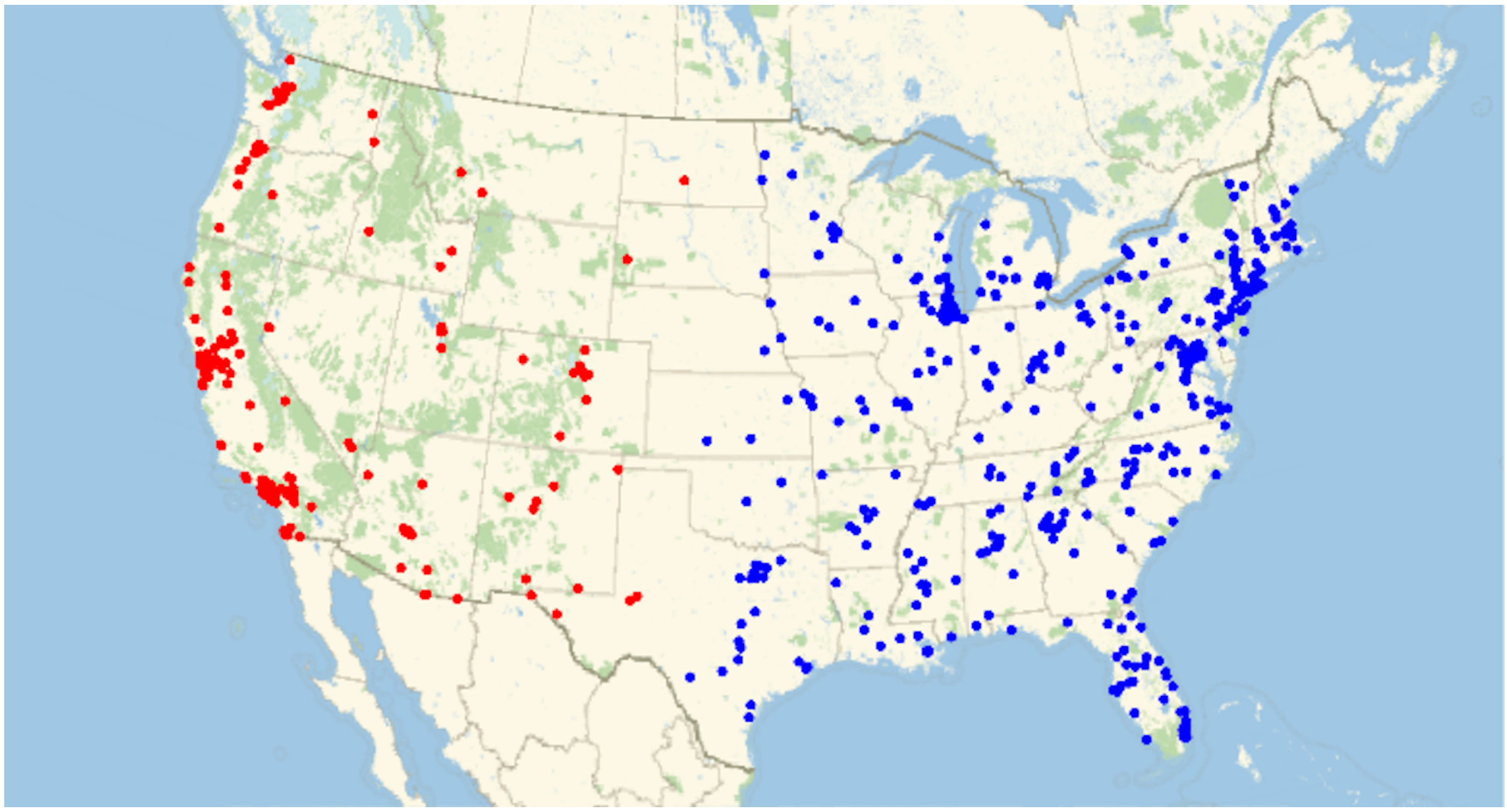

Fig. 5. k-means clusters for k = 2

For k = 3 and 4, k-means clustering provides further geographically identifiable and coherent regions as clusters as shown in Fig. 6 and Fig. 7 respectively:

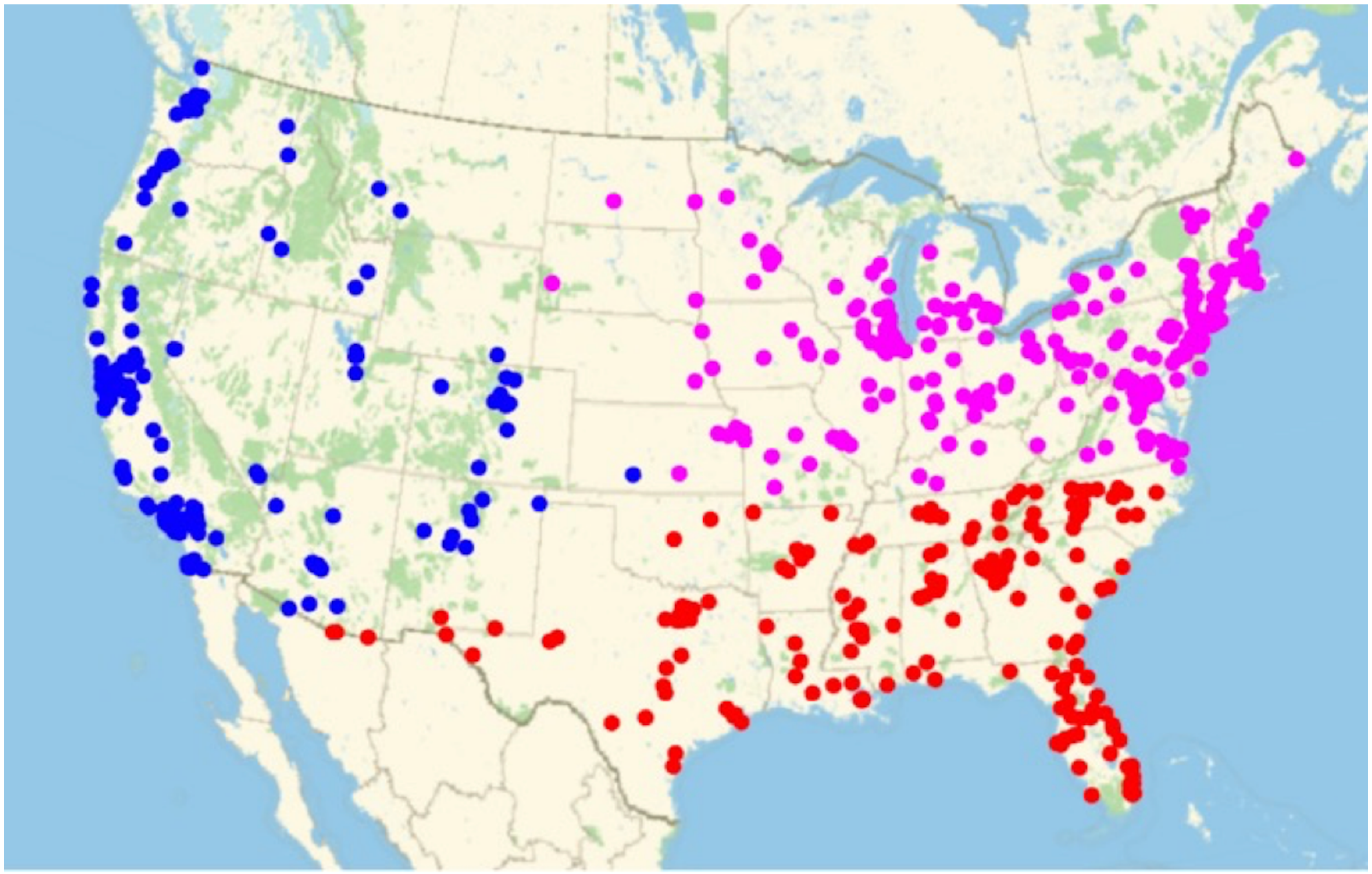

Fig. 6. k-means clusters for k = 3

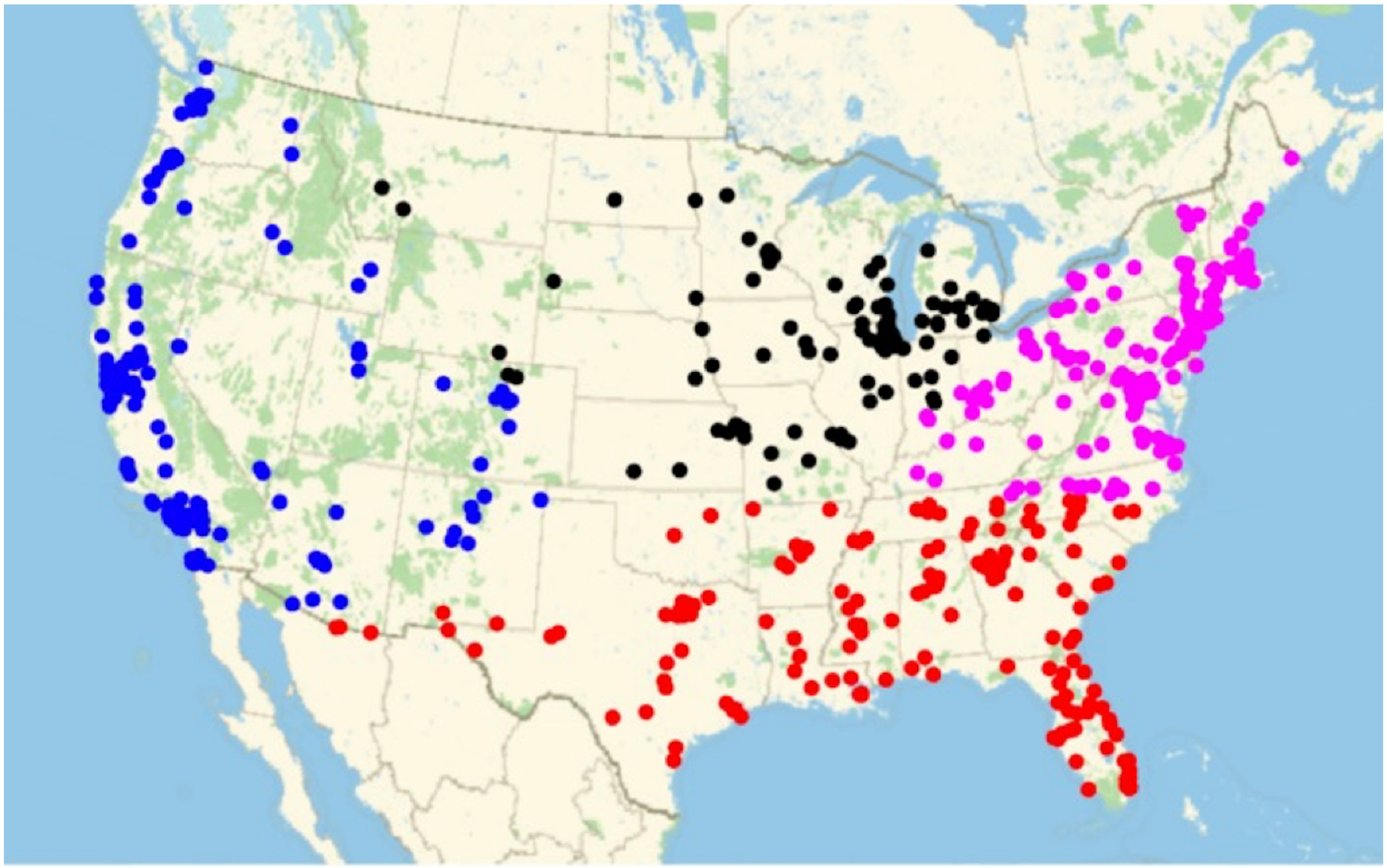

Fig. 7. k-means clusters for k = 4

k-means clustering seems to lead to reasonably well-defined geographic regions - the west, mid-west, east coast and southern US. The geographic nature of the clusters satisfies Bacher et al.'s criterion of *interpretability*: "It must be possible to give the clusters substantive meaningful names." (Bacher et al., 2022, p. 345)

*3.3 County level political violence*

We observe that locations of political violence are more densely distributed along the coasts, both east and west, and much more sparsely distributed in the center of the US, where there are regions with apparently no locations of political violence at all. To specify more precisely what we mean by "region" we could, for example, divide the continental US into grid regions (ref. Diggle, 2003). Doing so requires us to specify the size of the grid regions. Instead, we use a naturally occurring division into administrative regions, namely counties, of which there are approximately 3143 in the US. Instead of counting, for each county, the number of unique *locations* of political violence in that county, we instead count the number of *incidents* of political violence occurring in that county.

There are 393 counties with one or more incidents of political violence. This is 13.0 % of all US counties. Additionally, only 6.1 % of all US counties had 2 or more incidents of political violence, and only 2.7% had 5 or more such incidents.
For each county with at least one recorded incidence of political violence we counted the number of such incidents in that county. The distribution of such counts is highly skewed, with a heavy tail (skewness = 8.66, kurtosis = 97.46, maximum = 167) as shown in Fig. 8.

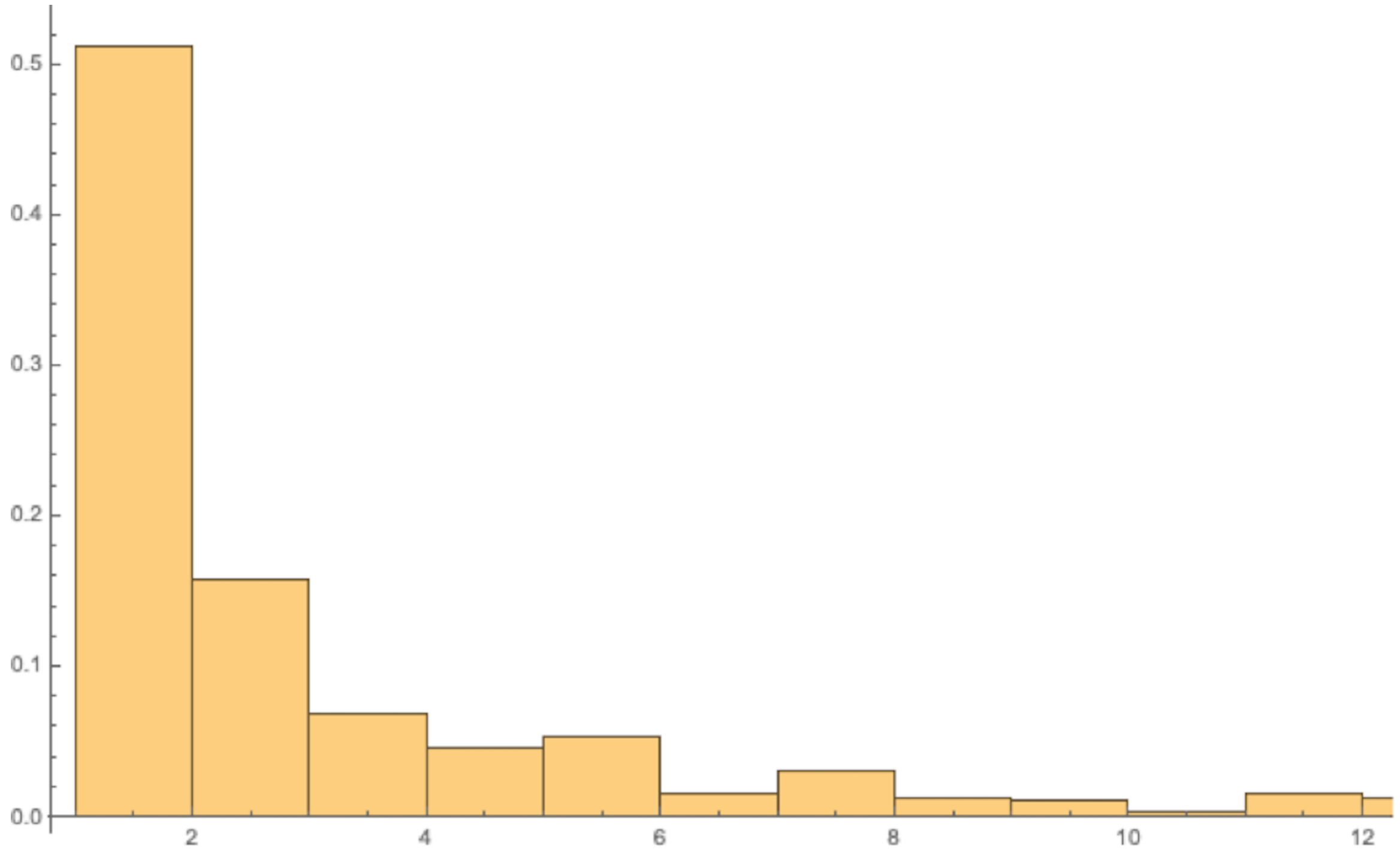

Fig. 8. Distribution of counts of incidents of political violence per county

The mean of county counts is 4.61 and the standard deviation is 12.22. We chose to examine in some detail those counties with a z-score of their counts of incidents of political violence 3 or greater. Were the count data normally distributed we would expect only about 0.13% of cases to have z-score 3 or greater. However, we find 6 counties, or 1.53% of all cases, with z-score 3 or greater. All these counties have more than 40 recorded incidents of political violence in a 4-year span. They are shown below in Table:

**Table 1**

Top 6 counties with highest counts of incidents of political violence

| County | Incidents of political violence | z-score |
|---|---|---|
| Multnomah County, Oregon | 167 | 13.29 |
| Los Angeles County, California | 102 | 7.97 |
| New York County (Manhattan), New York | 88 | 6.82 |
| Kings County (Brooklyn), New York | 54 | 4.04 |
| King County, Washington | 53 | 3.96 |
| District of Columbia, District of Columbia | 44 | 3.22 |

We observe Multnomah County, Oregon, as an extreme outlier with 167 recorded incidents of political violence. These instances of political violence occurred in 10 different locations in Multnomah County, with 65 cases occurring at the location with latitude 45.5134, longitude -122.681, which Google maps identifies as the OHSU Market Square building at 544 SW Market St, Portland, OR 97201.

The ACLED database contains the news source and a news commentary for each recorded act of political violence. The news commentary from the ACLED data file related to violent incidents at or near this location - latitude 45.5134, longitude -122.681 - were extracted and a word cloud generated from Python's WordCloud function, as shown in Fig. 9.

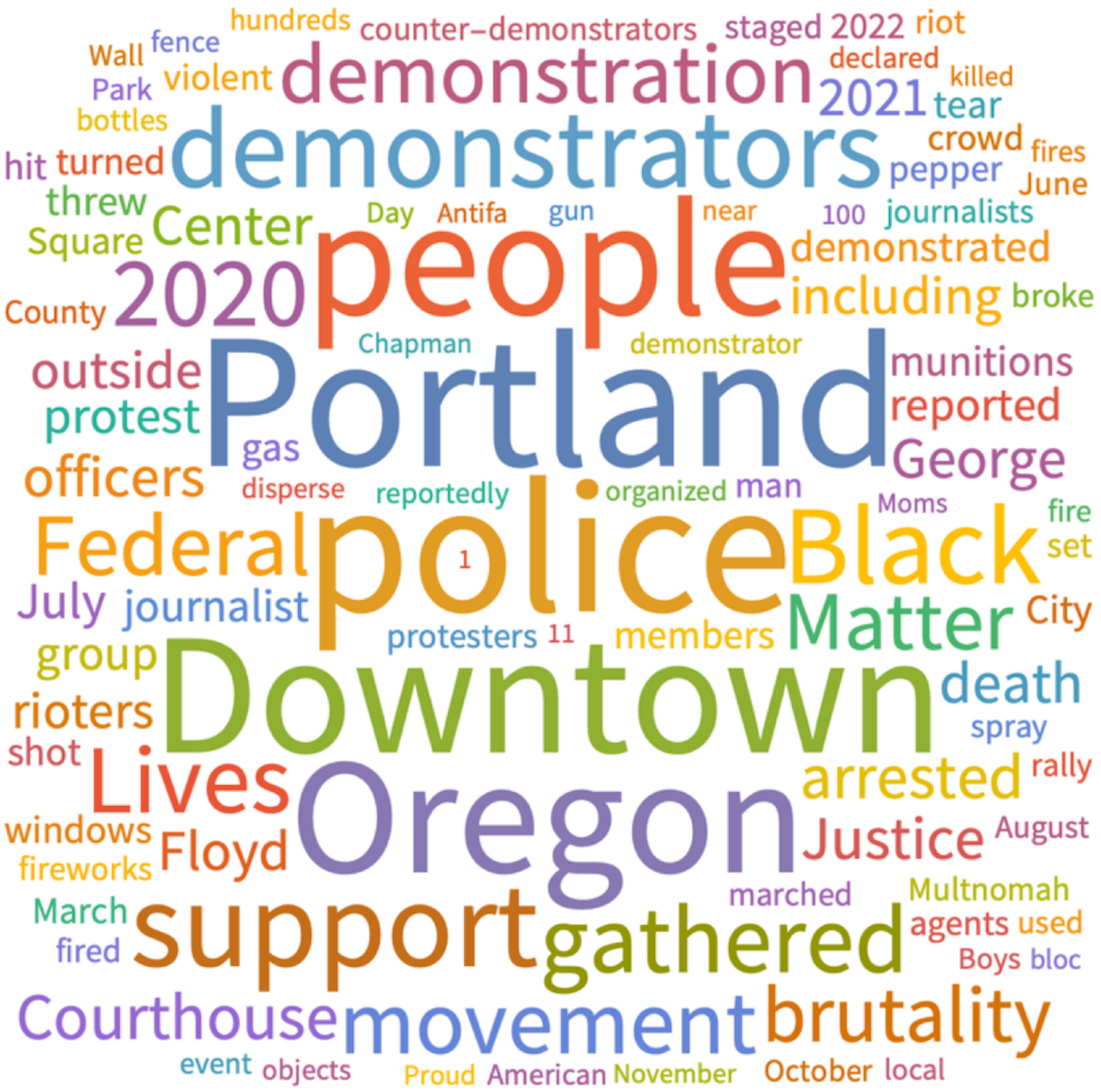

Fig. 9 Word cloud for commentary on violent incidents at latitude 45.5134, longitude -122.681

The word cloud and written commentary at the time, indicate Multnomah County was, and still may be, suffering gross racial inequalities related to health, education, housing and profiling, among other things. Multnomah County Board of County Commissioners declared racism a public health crisis on April 8, 2021, to address health inequities and disparities in access to quality education, employment, housing, and healthcare. Timur Ender, Multnomah County Public Health Advisory Board: “Violence is not random. It’s driven by policy decisions, from a lack of sidewalks to inequities in the justice system.” (Multnomah County Communications Office, 2021). Bates et al., 2014, also comment on persistent racial inequities in Multnomah county:

"African-Americans in Multnomah County continue to live with the effects of racialized policies, practices, and decision-making. The stress of racism has a profound impact on health and wellness, as do other social determinants of health, such as ongoing discrimination in housing, school discipline, and racial profiling by police. Multiple systems of inequity and institutional racism results in over-representation in punitive systems like juvenile justice and child welfare, and underrepresentation in systems that advance opportunities, like higher education and civil services employment. " (p. 2)

The ACLED data contains a daily time stamp - recording year, month and day - for each recorded act of political violence. This allows us to relatively easily construct and visualize time series counts of acts of political basis by day or by month. A plot of the monthly time series of all violent demonstrations (not just those in Multnomah county) is shown in Fig. 10 below, together with the mean (dashed black) and 1.96 standard deviations above the mean (dashed red):

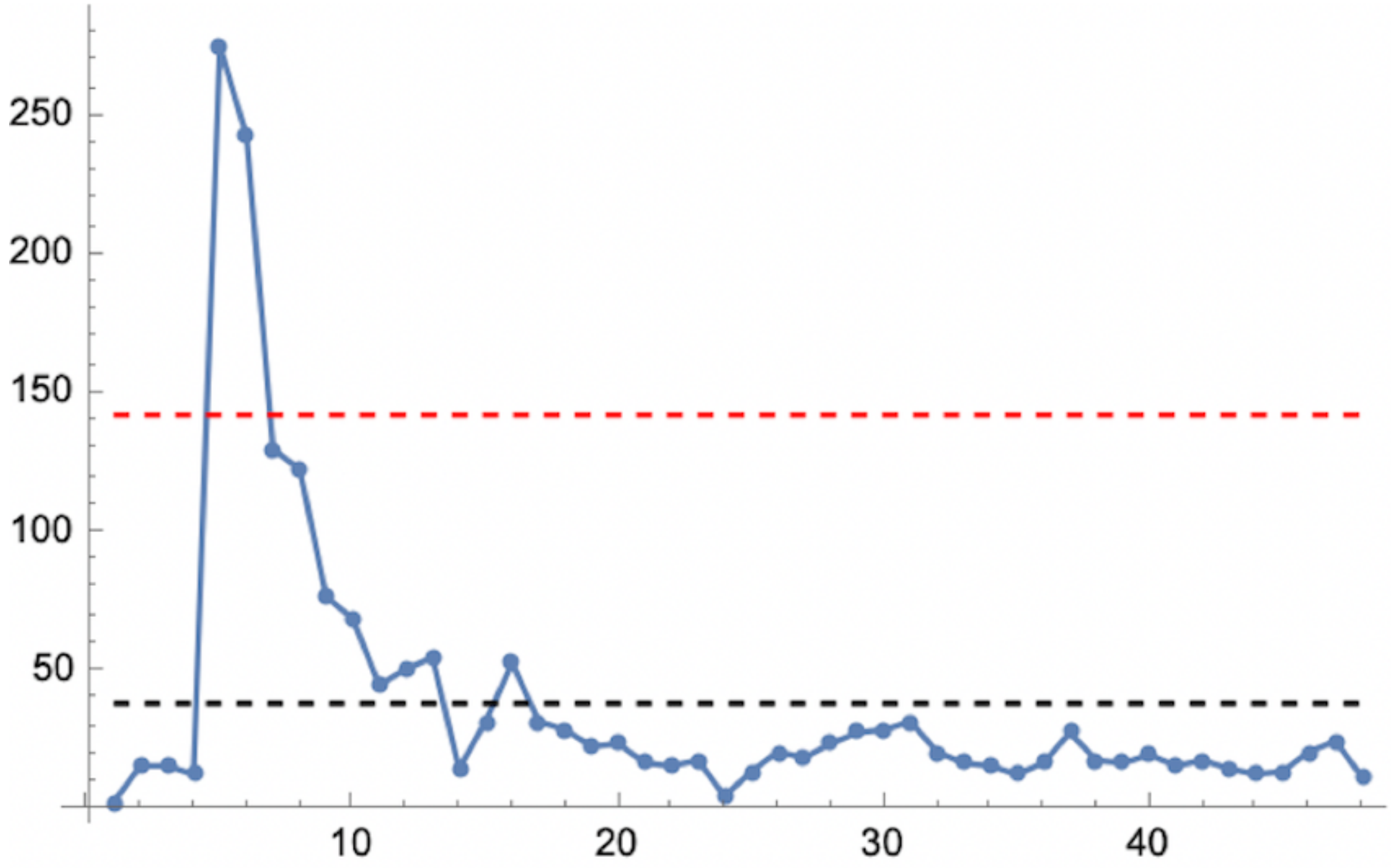

Fig. 10. Monthly time series of all violent demonstrations with the mean (dashed black) and 1.96 standard deviations above the mean (dashed red)

We note the extreme outliers in May and June 2020. The word cloud strongly suggests those violent demonstrations may have been largely connected with the death of George Floyd on May 25, 2020.

### *3.4 Political violence at specific locations*

Here we focus on specific locations (latitude and longitude) at which there are high levels of recorded political violence. As for the counties data the distribution of number of incidents of political violence at specific locations is heavily skewed, with a heavy tail (skewness = 7.33, kurtosis = 74.3, maximum = 87), as shown in Fig. 11.

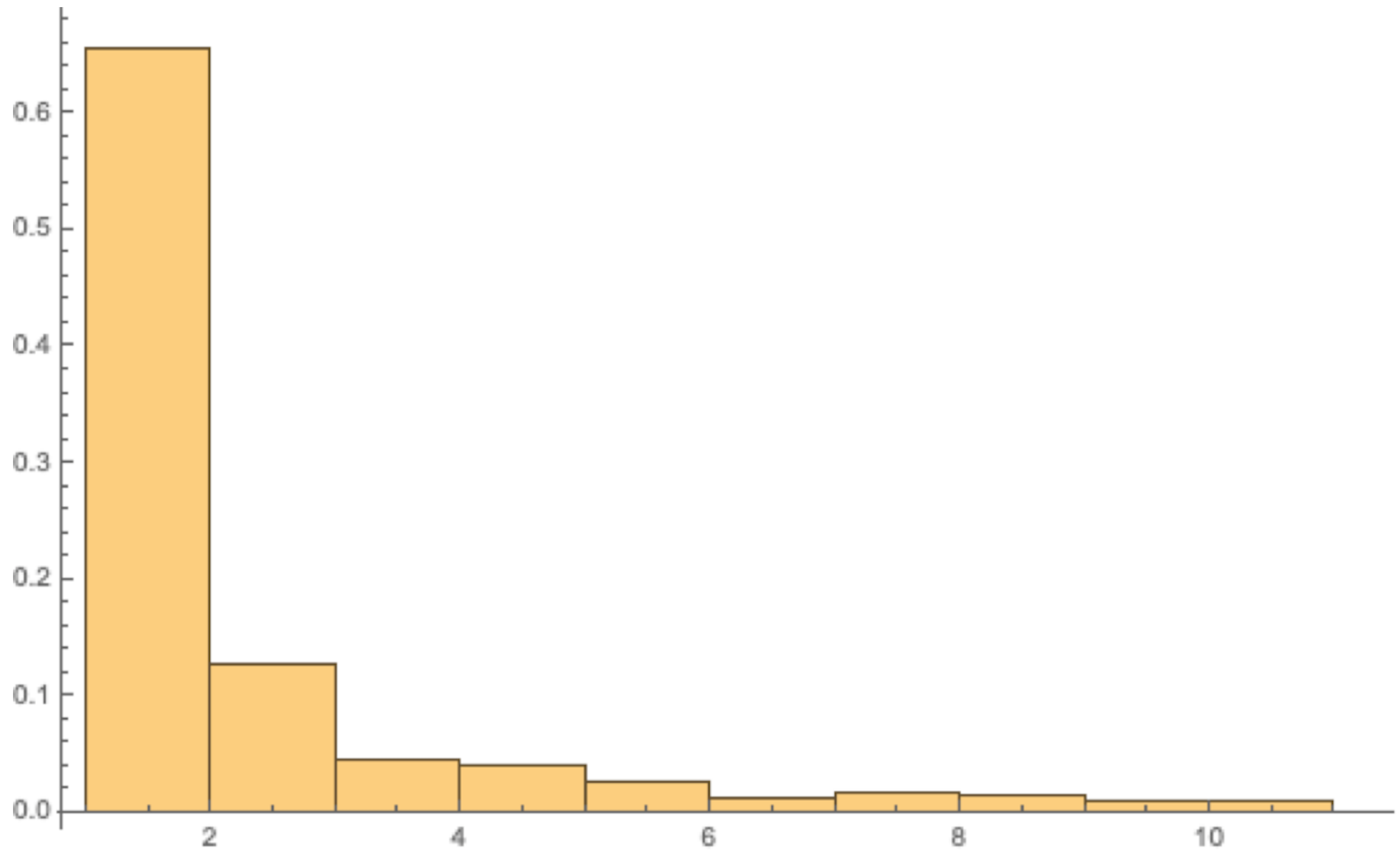

Fig. 11. Distribution of counts of incidents of political violence per location

As for counties, we focus on those locations at which the number of incidents of political violence has z-score 3 or greater. These locations are shown in Table 2:

**Table 2**
List of locations with high number of incidents of political violence

| latitude | longitude | Incidents | City & state | General description |
|---|---|---|---|---|
| 40.7834 | -73.9663 | 87 | Central Park NY | 86th St Transverse & Central Park W |
| 45.5134 | -122.6809 | 65 | Portland OR | SW Clay St & SW 5th Ave |
| 40.6904 | -73.9862 | 54 | Broooklyn NY | Hoyt & Livingston |
| 47.6056 | -122.333 | 45 | Seattle WA | 4th Ave & Washington |
| 33.7559 | -84.3898 | 37 | Atalanta GA | Broad St NW & Walton St NW |
| 45.5234 | -122.6762 | 31 | Portland OR | NW 6th AVE & W Burnside St |
| 44.9833 | -93.2666 | 27 | Minneapolis MN | S 1st St & Hennepin Ave |
| 45.5761 | -122.7246 | 26 | Portland OR | N Haven Ave & N Yale St |
| 34.0693 | -118.3118 | 25 | Los Angeles CA | W 3rd St & S St Andrews Place |
| 38.5799 | -121.4962 | 24 | Sacramento CA | K St & 8th St |

Geographically this seems to be a somewhat scattered collection of locations. We focus on the location (40.7834, -73.9663) with the highest number of recorded incidents of political violence (87) in Central Park NYC. This is a central location in Manhattan, a city with a very large population. We might expect therefore a variety of reasons why incidents of political violence occurred at this precise location. A word cloud was extracted from the news source and news commentary in the ACLED database, as shown in Fig. 12.

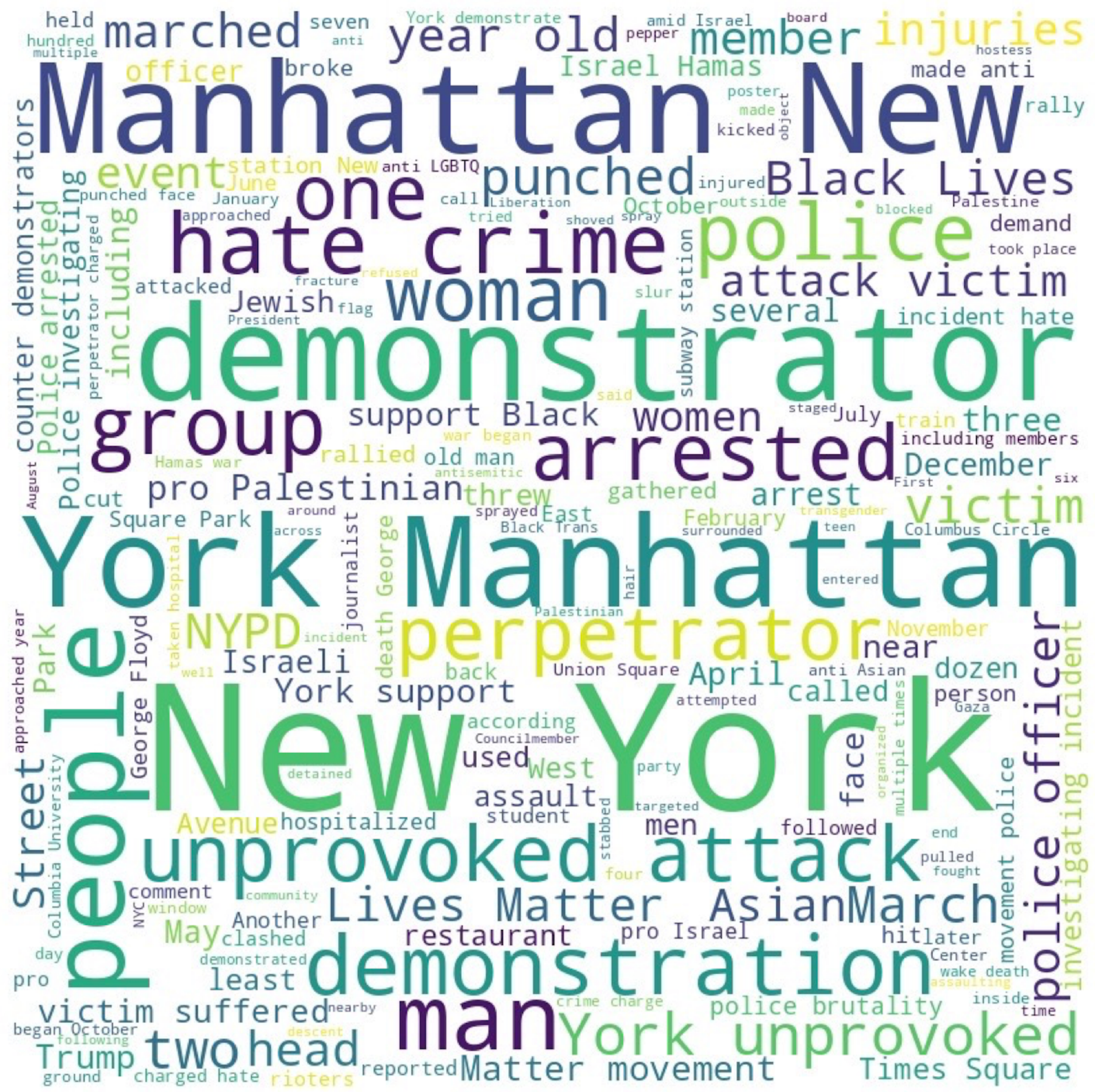

Fig. 12 Word cloud for commentary on violent incidents at latitude 40.7834, longitude -73.9663

We find a variety of words and phrases from the news commentary related to incidents of political violence, including comments on race, ethnicity, religion, sexual preference, political allegiance, and perceived injustices. As one might expect, there is no one clear theme to these 87 incidents of political violence. However, these incidents do, in fact, involve violence - they are not simply demonstrations. On the face of it these would seem to be demonstrations that have turned violent. If that is so, it is an indicator of an underlying level of upsetness of those involved.  For example, from the ACLED news commentary we have the following description:

> " On 19 March 2023, roughly 40 people, including Proud Boys and Orthodox Rabbis, gathered at the LGBTQ+ Community Center in New York - Manhattan (New York) to demonstrate against a Drag Story Hour being hosted at the center which featured State

Attorney General Letitia James. Some demonstrators also waved pro-Trump signs. Nearby, more than 100 counter-demonstrators, including New York City Council Member Erik Bottcher (D), rallied in support of the LGBTQ+ community. A masked anti-event demonstrator was arrested for allegedly assaulting someone, causing them to bleed, and punching and grabbing at cameras. Some reports indicate that there were additional clashes between pro- and anti-event demonstrators, including Proud Boys. "

*3.5 Scaling by population*

It has not escaped our attention that the number of incidents of political violence per county may be correlated with the county population – in other words, higher population may lead to a greater number of incidents of political violence. To address this we calculated for each county in which there was at least one incident of political violence the ratio of number of recorded incidents of political violence per 1000 head of population. We refer to this simply as "ratio" in the table below. The estimate of a county's population was obtained by averaging estimates for the years 2020 through 2023 (ref. United States Census Bureau).

**Table 3**
Ratio of number of recorded incidents of political violence per 1000 head of population, and number of recorded incidents, sorted by highest to lowest ratio.

| County | ratio | # incidents |
|---|---|---|
| Kiowa County, Kansas, United States | 0.414809 | 1 |
| Hudspeth County, Texas, United States | 0.299895 | 1 |
| Union County, New Mexico, United States | 0.248247 | 1 |
| Multnomah County, Oregon, United States | 0.220829 | 167 |
| Rio Blanco County, Colorado, United States | 0.153374 | 1 |
| Refugio County, Texas, United States | 0.149499 | 1 |
| Clear Creek County, Colorado, United States | 0.107055 | 1 |
| Jefferson County, Montana, United States | 0.0792252 | 1 |
| Winn Parish, Louisiana, United States | 0.0745156 | 1 |
| Screven County, Georgia, United States | 0.0710694 | 1 |

Apart from Multnomah County, we see that a number of counties with only 1 incident of political violence rise to the top in terms of the calculated ratio, due to their relatively small population.

For the top 10 counties based on highest number of incidents of political violence we checked their rank according to the above calculated ratio:

**Table 4**
Rank, by calculated ratio, for the 10 counties with the highest number of incidents of political violence

| County | Ratio rank |
|---|---|
| Multnomah County, Oregon, United States | 4 |
| Los Angeles County, California, United States | 144 |
| New York County Manhattan, New York, United States | 19 |
| King County, Washington, United States | 64 |
| Kings County Brooklyn, New York, United States | 90 |
| District of Columbia, District of Columbia, United States | 11 |
| Fulton County, Georgia, United States | 35 |
| Hennepin County, Minnesota, United States | 52 |
| Alameda County, California, United States | 97 |
| Sacramento County, California, United States | 100 |

We conclude there is no clear connection between number of incidents of political violence per county and the county population.

*3.6 Nearest neighbor classification of fatalities*

The number of fatalities at a US incident of political violence ranges from 0 through 22, with a median of 0 and a mean of 0.184. Approximately 89% of US violent incidents have no associated fatalities.

Imagine law enforcement intelligence forecasts a potentially violent demonstration at a future date (year, month, day) and a specific location (latitude, longitude). Can we use existing data to classify this future event as one that will, or will not, have associated fatalities? And how accurate can we be with such a classification?

To carry out classification with k-nearest neighbors, we label a space-time event (year, month, day, latitude, longitude) "0" if it has 0 associated fatalities, and "1" if it has 1 or more associated fatalities. So, a typical data point is of the form (year, month, day, latitude, longitude, fatality) where "fatality" is the 0-1 fatality label.

**Distance measure for space-time events**: We measure distance between two space-time events (year1, month1, day1, latitude1, longitude1) and (year2, month2, day2, latitude2, longitude2) as the unsigned number of days between (year1, month1, day1) and (year2, month2, day2) plus geographic distance between (latitude1, longitude1) and (latitude2, longitude2).

We split our US data into training data 2020-2022 and test data 2023-2024. For a given 2023-2024 data point (year, month, day, latitude, longitude, fatality) and a given odd k, we find the k nearest neighbors to the space-time data point (year, month, day, latitude, longitude) in the 2020-2022 space-time data, using the distance measure described above. Each of these k nearest neighbors has a fatality label, either 0 or 1, and we take the majority of 0s and 1s to be the predicted label of the 2023-2024 data point (year, month, day, latitude, longitude). We return a value of TRUE if the fatality label of the 2023-2024 event is the same as the predicted label, and FALSE otherwise. We do this for each 2023-2024 event and record the % TRUE cases as the % accuracy of the k-nearest neighbor classification, based on the 2020-2022 training data and the 2023-2024 test data. Results for odd k from 1 through 31 are shown in the Fig. 13 below:

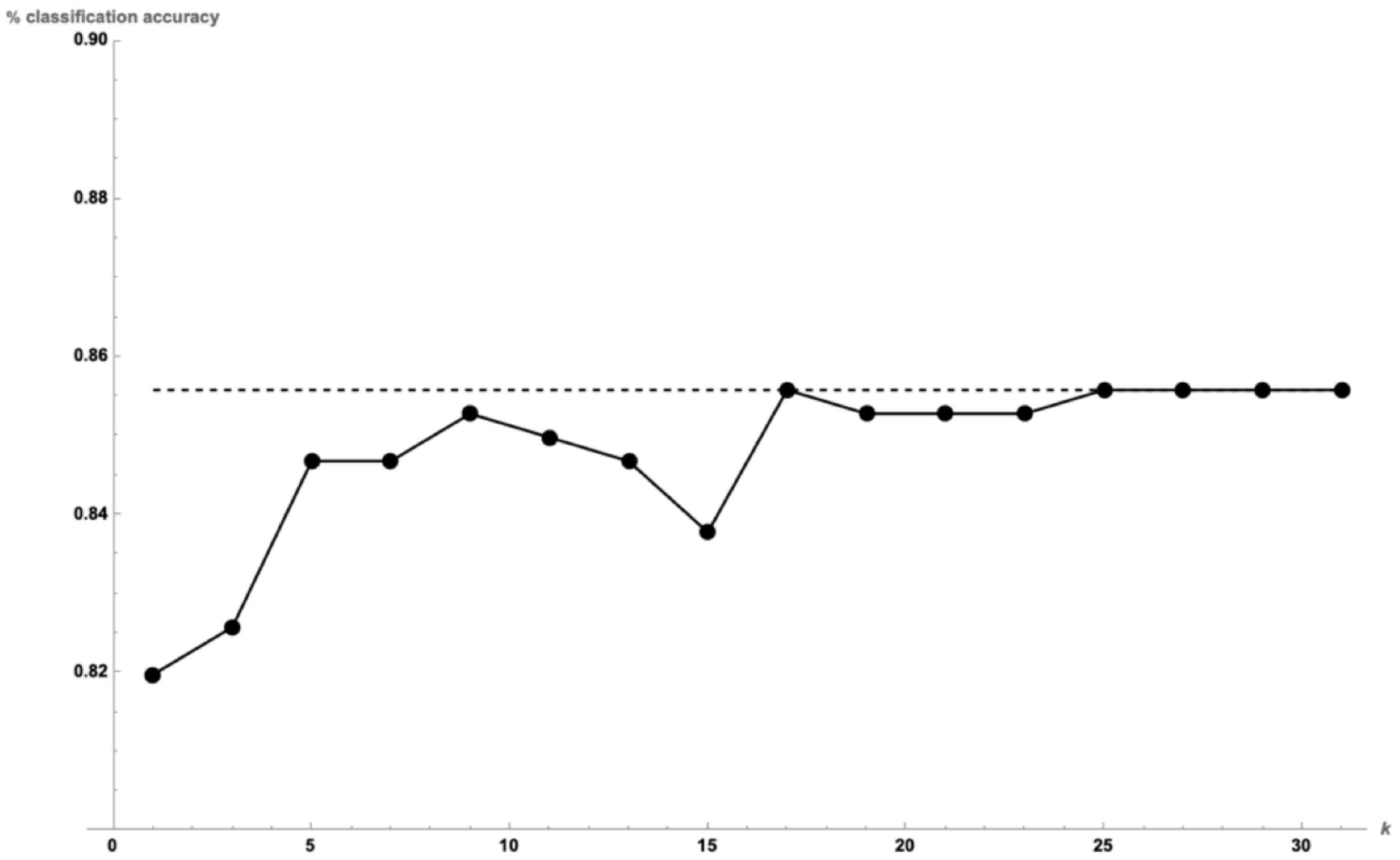

Fig. 13. Percentage classification accuracy versus odd k

We observe approximately 85% classification accuracy for k = 9, and approximately 86% classification accuracy for k = 17, with little to no higher percentage accuracy for larger values of k. This k-nearest neighbor classification procedure can be readily updated account for new training data. The sizes of the violent events data are not so large as to create a scale problem for the lazy k-nearest neighbor algorithm and searching for nearest neighbors is an easily parallelized algorithm, contributing significantly to speeding up computation.

## 4. Fields of upsetness: a model of public outrage and violent action

A field notion of upsetness leading to political violence by individuals and groups arises naturally, and somewhat compellingly, from a study of conditions leading to US counties that are outliers for the number of incidents of political violence. Detailed report and commentary on the political and social situations in these locations indicates to us an ambient environment in which many people are upset for extended periods of time. This leads us to imagine a field of upsetness, where our general understanding and explication of a field notion of political violence is in accord with the writing of Martin, 2003. Our field notion postulates a *field of upsetnesss* by individuals in a specific region to the point where a trigger event can act to promote political violence. Purported characteristics of a field of upsetness are:

- A field explains changes in the states of some people but need not appeal to changes in states of other people (as a "cause" of change, for example).
- These changes in state involve an interaction between a field of upsetness and the existing states of people.
- Individual people have particular attributes that make them susceptible to a field effect

- A field of upsetness without people is only a potential for creation of a force, without any existent force.

People influence each other through messaging. In the past this would have been through personal contact, newspapers, TV news, or public speeches by politicians, or other community gatherings. With the advent of social media and cell phones the possibilities for messaging have expanded enormously. As many individuals send and receive messages, they build a field of views and opinions. When those views and opinions lead the individuals to being upset, angry or insecure we claim that the messaging has established a field of upsetness. Kleinfeld, for example, argues that "... while social factors may have created the conditions, politicians have the match to light the tinder." (Kleinfeld , 2021, p. 173.) This upsetness field, we claim is as real in the social sphere, as an electromagnetic field, or a gravitational field, in physics. We discuss a significant example of a field of upsetnesss below.

**George Floyd and Black Lives Matter protests**

Our time series plot of monthly incidents of political violence shows the months of May and June 2020 to be major outliers (ref. figure 10). For Multnomah County, with the highest count of political violence of any US county, there were 23 incidents of political violence in July, 202 and 33 such incidents in August 2020. A word cloud of news reports for those two months highlighted the phrases: Black Lives Matter, demonstration, police brutality, tear gas, George Floyd, riot, turned violent.

This indicates to us that the high count of political violence occurred in a background of heightened racial inequity, and the specific trigger of the killing of George Floyd (May 2020) inflamed that perceived inequity into specific acts of political violence.

On May 25, 2020, Minneapolis police officers arrested George Perry Floyd Jr., a 46-year-old father of five and grandfather of two for allegedly buying cigarettes with a counterfeit $20 bill. Derek Chauvin, a white police officer, who was one of the four officers (Derek Chauvin, Thomas Lane, Tou Thao and J. Alexander Kueng) at the scene had arrested George Floyd by putting him in handcuffs and then pinned him on the ground for more than 9 minutes. All three officers held Mr. Floyd in a position that restricted his breathing. Emergency responders arrived, and medics loaded him into an ambulance. He was pronounced dead that night.

All these incidents were recorded on bystanders' mobile phones. A disturbing video from one of the bystanders incited large protests against police brutality and systemic racism in Minneapolis and across the United States in the months that followed, leading to a racial justice movement not seen since the civil rights protests of the 1960s. The National Guard was activated in at least 21 states, and cities announced curfews as protesters filled the streets for demonstrations that sometimes turned destructive. Law enforcement was criticized for responding to the protests — a majority of which were peaceful — with force, by spraying tear gas and shooting rubber bullets at protesters and conducting mass arrests. Images from representative incidents are shown in Fig. 14.

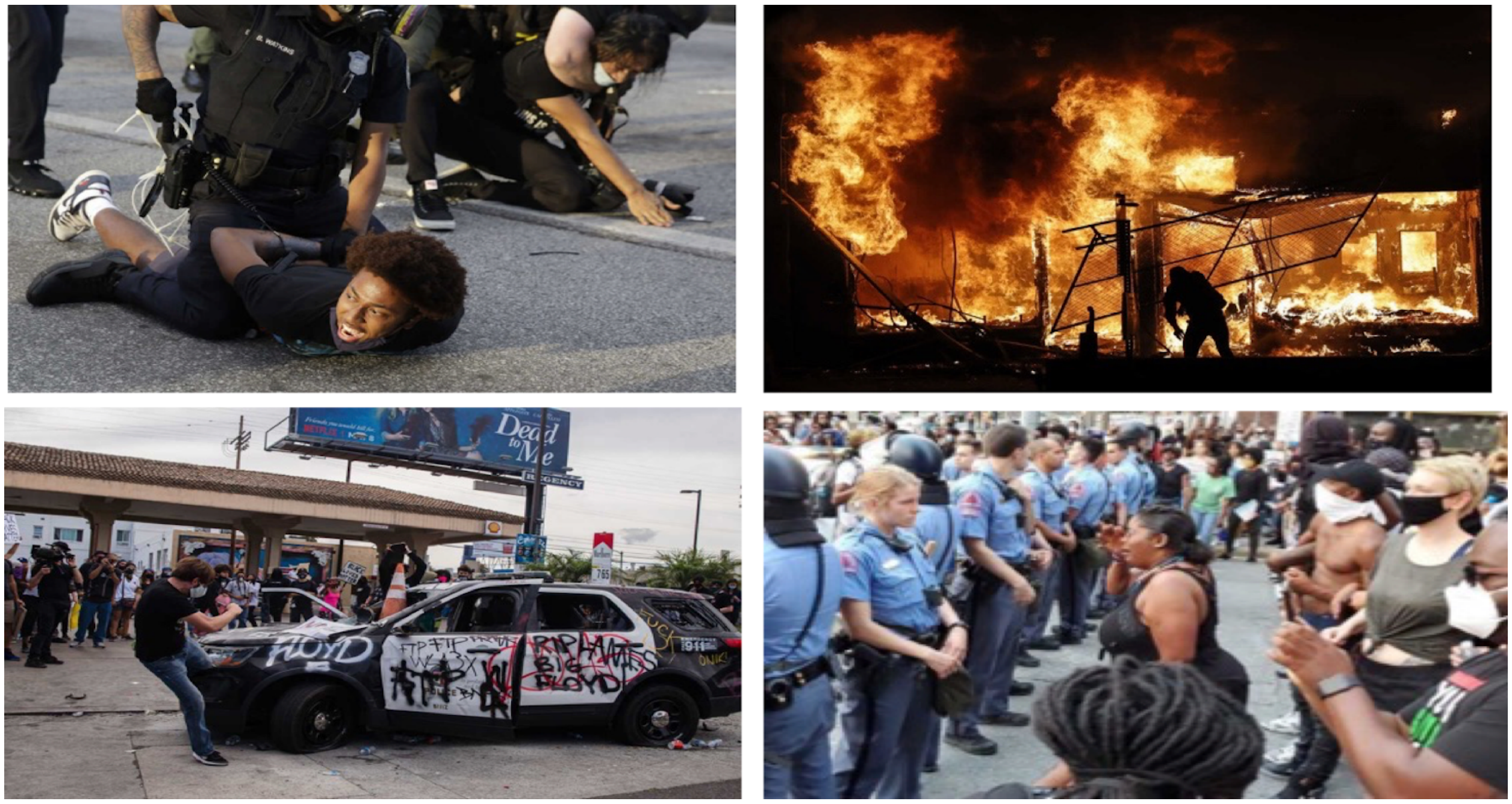

Fig. 14. George Floyd killing outraged people already upset at treatment by police officers.

*Social media*

The role of social media in stimulating political violence has been examine by several authors: Alava et al., 2017, examine the role of social media on the radicalization of youth leading to violent extremism in Europe, North America, Latin-America and the Caribbean (pp. 13-26); Pauwels & Hardyns, 2018, examine connections between self-reported political aggression and social media exposure:

> "Endorsement for extremism is related to self-reported political aggression, but the effect of endorsement increases by levels of active social media exposure." (p. 51).

Schleimer et al., 2024 report on the association between many strong social connections and the likelihood of both endorsing, and being personally willing to use, political violence:

> " ... individuals who reported 50 + strong social connections (4.2% of respondents) were also more likely to endorse political violence in specific situations and to be personally willing to use political violence." (p. 10)

In relation to social media influence, Kleinfeld (2021, p. 160) states:

> "According to the National Consortium for the Study of Terrorism and Responses to Terrorism (START), which maintains the Global Terrorism Database, most political violence in the United States is committed by people who do not belong to any formal organization. Instead, ideas that were once confined to fringe groups now appear in the mainstream media. White-supremacist ideas, militia fashion, and conspiracy theories spread via gaming websites, YouTube channels, and blogs, while a slippery language of

> memes, slang, and jokes blurs the line between posturing and provoking violence, normalizing radical ideologies and activities."

The role of social media in inflaming passions, leading to riots, not only in the United States, but also in other jurisdictions, is spelled out with considerable clarity by Dame Melanie Dawes, Chief Executive of Ofcom, in a letter to the UK Secretary of State, October 2024, in relation to the 2024 UK riots which originated in Southport and spread to over 20 other UK locations:

> **Illegal content and disinformation spread widely and quickly online following the attack** (our emphasis). While not all platforms experienced significant levels of illegal and harmful material, others have confirmed to us they were dealing with high volumes, reaching the tens of thousands of posts in some cases. **Misinformation appeared online almost immediately after the attacks, some of it appearing to have malicious intent and seeking to influence public opinion and reaction** (our emphasis). We have heard how some platforms were used to spread hatred, provoke violence targeting racial and religious groups, and encourage others to attack and set fire to mosques and asylum accommodation. Accounts (including some with over 100,000 followers) falsely stated that the attacker was a Muslim asylum seeker and shared unverified claims about his political and ideological views. Posts about the Southport incident and subsequent events from high-profile accounts reached millions of users, **demonstrating the role that virality and algorithmic recommendations can play in driving divisive narratives in a crisis period** (our emphasis).
>
> **There was a clear connection between online activity and violent disorder seen on UK streets** (our emphasis). Some major platforms were used to post and disseminate material calling for violent action in Southport and in other towns and cities. One analyst told us that calls for demonstrations targeting a local mosque were circulating in private groups online within two hours of the vigil for the victims of the attack. **The evidence we have seen indicates to us that some messaging services hosted closed groups comprising thousands of users in some cases. Some of these groups disseminated material encouraging racial and religious hatred, and provoking violence and damage to people and property, including by identifying potential targets for damage or arson** (our emphasis). (Dawes, 2024)

The role of social media in exacerbating existing grievances and provoking acts of political violence is also described clearly in a Reuter's (2024) report relating to the same July 2024 UK riots.

## 5. Conclusions

While political violence in the United States is on the increase, the geographic distribution of incidents of political violence is very uneven, as data from the Armed Conflict Location & Event Data Project (ACLED) shows. A geographic analysis of locations of political violence in the continental United State (January 2020 through July 2024) indicates locations of political violence are not uniformly random or Poisson random. Locations of political violence generally fall into large geographic clusters: for example, coastal regions generally had more incidents of political violence than did regions in the center of the US.

At the level of US county, we find the number of incidents of political violence per county is highly skewed, with most counties (87%) having no recorded incidents of political violence. Similarly, the number of incidents of political violence at a specific latitude and longitude is also highly skewed. We found 6 counties (1.53% of all cases) with z-score of counts of political violence 3 or greater. All these counties had more than 40 recorded incidents of political violence in a 4-year span. A time series analysis of incidents of political violence also exposes certain critical events as major promoters of political violence. A field notion of upsetness leading to political violence by individuals and groups arises naturally, and somewhat compellingly, from a study of conditions leading to US counties that are outliers for the number of incidents of political violence.
A detailed examination of news reports and commentaries from the ACLED database revealed compelling social and political statements supporting an underlying environment of multiple systems of inequity and institutional racism. Certain critical events can spark political violence in an already upset community. These events usually occur in a background of online "chattering", distributed and shared via social media. The voices given to people, including politicians and lawmakers, by social media have created a phenomenon in society where it is increasingly easy to spread false and misleading statements as true and factual. This intentionally spread misinformation creates ambient states, or fields, of upsetness leading potentially to grievance and political violence, and these fields of upsetness are strengthened by information bubbles, in which there is little to no voiced opposition to commonly held views and opinions.
An issue that is somewhat puzzling for which we currently have no answer is: How does the empirically observed highly uneven geographic distribution of incidents of political violence, relate to the fact that for communication purposes, geographic distance is minimized because of the use cell-phones and social media?